\documentclass[aps,prl,preprint]{revtex4-1}

\usepackage{graphicx}  % needed for figures
\usepackage{amsmath} %needed for math
\usepackage{dcolumn}   % needed for some tables
\usepackage{bm}        % for math
\usepackage{amssymb}   % for math
\usepackage{epstopdf}
\usepackage{xcolor} %for coloured text
\usepackage[normalem]{ulem}

\def \lf {L$_\phi$~}
\def \a {$\alpha$~}
\def \Vtg {V$_{TG}$}
\def \Vbg {V$_{BG}$}

\def \BSTS {BiSbTe$_{1.25}$Se$_{1.75}$}

\begin{document}

\title{Spatially varying electronic dephasing in three dimensional topological insulators} 

\author{Abhishek Banerjee$^1$, Ananthesh Sundaresh$^1$, R. Ganesan$^1$, 
 and P. S. Anil Kumar$^1$}
\affiliation{\small{$^1$Department of Physics, Indian Institute of Science, 
Bengaluru 560012, India }}

%\date{\today}
\begin{abstract}
Information processing devices operating in the quantum mechanical regime strongly rely on the quantum coherence of charge carriers. Studies of electronic dephasing in conventional metallic and semiconductor systems have not only paved the way towards high coherence quantum electronics, but also led to fundamental new insights in condensed matter physics. In this work, we perform a spatially resolved study of electronic dephasing in three dimensional topological insulators by exploiting an edge versus surface contacted measurement scheme. Unlike conventional two dimensional systems that are characterized by a single dephasing mechanism, we find that dephasing in our samples evolves from a variable-range-hopping type mechanism on the sample surface to a Nyquist type electron-electron interaction mechanism in the sub-surface layers. This is confirmed independently by the temperature and chemical potential dependence of the dephasing length, and gate dependent suppression/enhancement of the weak anti-localization effect. Our devices are fabricated using bulk insulating topological insulator \BSTS~capped with hexagonal-Boron Nitride in an inert environment, ruling out any extrinsic effects and confirming the topological surface state origin of our results. Our work introduces the idea of spatially resolved electronic dephasing and reveals a new regime of coherent transport in perhaps the most important topological insulator discovered so far. Our edge-vs-surface scheme may be applied to dephasing studies in a wide class of 2D materials. 
 
\end{abstract}

\maketitle
\subsection{Introduction}

Topological insulators are materials characterized by an insulating bulk and conducting topological surface states(TSS)~\cite{TIreview1, TIreview2}. These electronic states are gapless and feature a spin-momentum locked structure that can give rise to unusual physics. The discovery of the 3D topological insulator phase in the Bi$_2$Se$_3$ material class sparked the birth of the first topological insulator with a simple band-structure~\cite{TIreview3, TIreview4, TIreview5} . However, it was quickly realized that Bi$_2$Se$_3$ material class suffered from unwanted bulk conduction due to a large concentration of defects, making them behave as weak-metals rather than true insulators, thereby masking the properties of the TSS. Soon after, compositional alloying was used to compensate the residual bulk conduction in materials of the Bi$_x$Sb$_{2-x}$Te$_y$Se$_{3-y}$ class~\cite{insulatingTI1, insulatingTI2, insulatingTI3} ushering in the second generation of topological insulators with dominant surface state transport. These materials spawned a panoply of phenomenal breakthroughs in condensed matter physics including the observation of the quantum anomalous Hall effect~\cite{QAHE1}, high temperature quantum Hall effects~\cite{QHE, QHTI2, QHTI3, QHTI4}, topological magneto-electric effects~\cite{axion_insulator} and Majorana edge modes in magnetic TI/superconductor hetero-interfaces~\cite{Majorana_TI_SC}.  

Yet, strong compositional disorder combined with the small band-gap of these materials leads to the formation of bulk and surface electron-hole puddles, as recently confirmed by scanning tunneling microscopy and optical conductivity experiments~\cite{puddle1, puddle2, puddle3, puddle4, puddle5}. Hybridization between topological surface states and defect state potentials have also been shown to generate strong resonances that can form diffusive impurity bands~\cite{xu2017disorder}. Compensation induced disorder is therefore expected to drastically affect the properties of topological surface state charge transport, especially those pertaining to quantum coherence. However, despite the enormous literature on bulk insulating TIs, quantum decoherence mechanisms operating in such systems remain poorly understood. An experimental investigation in this direction is impeded by several problems: i) Although the Bi$_x$Sb$_{2-x}$Te$_y$Se$_{3-y}$ class of materials feature reduced bulk carrier densities, to truly probe surface state transport, one requires vanishing bulk conductivity, at least at low sample temperatures. This condition is rarely satisfied by most TIs, even in the so-called bulk insulating regime. ii) These materials are prone to surface oxidation. To probe the effect of surface/bulk charge puddles, the effect of surface oxidation must be eliminated. This is often achieved by capping with Selenium/Tellurium. However these elements can readily react with the native BSTS surface and/or contribute to electrical transport~\cite{salehi2015stability}, making it difficult to separate the effect of charge puddles from that introduced by surface disorder from the capping material. iii) All TI devices studied till date feature a surface contact geometry, where the electrical contact metallizes only the sample surface. As we show subsequently, this scheme cannot measure the topological surface state wave-function in its entirety, and provides only part of the full picture.

In this work, we study phase coherent transport in fully tunable h-BN capped \BSTS\\devices prepared in an inert environment, with vanishing bulk-conductivity at low temperatures. Two different types of devices featuring surface contacted and edge contacted electrical leads are measured and contrasted. The surface contacted devices show evidence of strong surface scattering that induces a variable range hopping(VRH) type of decoherence of surface carriers. On the other hand, edge contacted devices probe not only the part of the TSS wave function localized on the top surface of the sample, but also that which extends into the bulk of the TI flake. Under these conditions, we obtain completely different transport characteristics indicating Nyquist type electron-electron interaction between carriers. To the best of our knowledge, such spatially varying electronic dephasing has not been observed in TIs or any other material system studied so far. We interpret these results as an effect of disorder induced delocalization of the topological surface state wave-function leading to poorer screening of impurity potentials on the topmost quintuple layers but stronger screening a few layers within the material bulk. 

\subsection{Experiment}
\subsubsection{{\bf Device fabrication and characterization}}
We choose the bulk insulating topological insulator \BSTS for our studies. Details of single crystal preparation and ARPES measurements are provided elsewhere~\cite{Banerjee16, lohani2017band}. Our samples feature vanishingly small bulk conductivities at samples temperatures $T <100K$, where electrical current is completely carried by surface states(see supplementary material section C). Samples prepared from the same crystal show quantum Hall effect~\cite{QHE_Banerjee} at relatively mild experimental conditions, attesting to the high quality of our devices.The fabrication of devices(surface or edge contacted) in this work starts with the preparation of h-BN/BSTS stacks assembled on Si/SiO$_2$ substrates using a dry Van der Waals transfer technique(See Ref.~\onlinecite{Banerjee16} and supplementary information section A). The entire exfoliation and transfer process is carried out within an Argon glove box with $<$0.1 ppm of Oxygen, ensuring that surface contamination effects due to oxidation and moisture are completely suppressed. Furthermore, the highly insulating and non-reactive nature of h-BN ensures that the capping layer does not provide an additional source of surface disorder. 

\begin{figure}[!t]
\includegraphics[width=1.\linewidth]{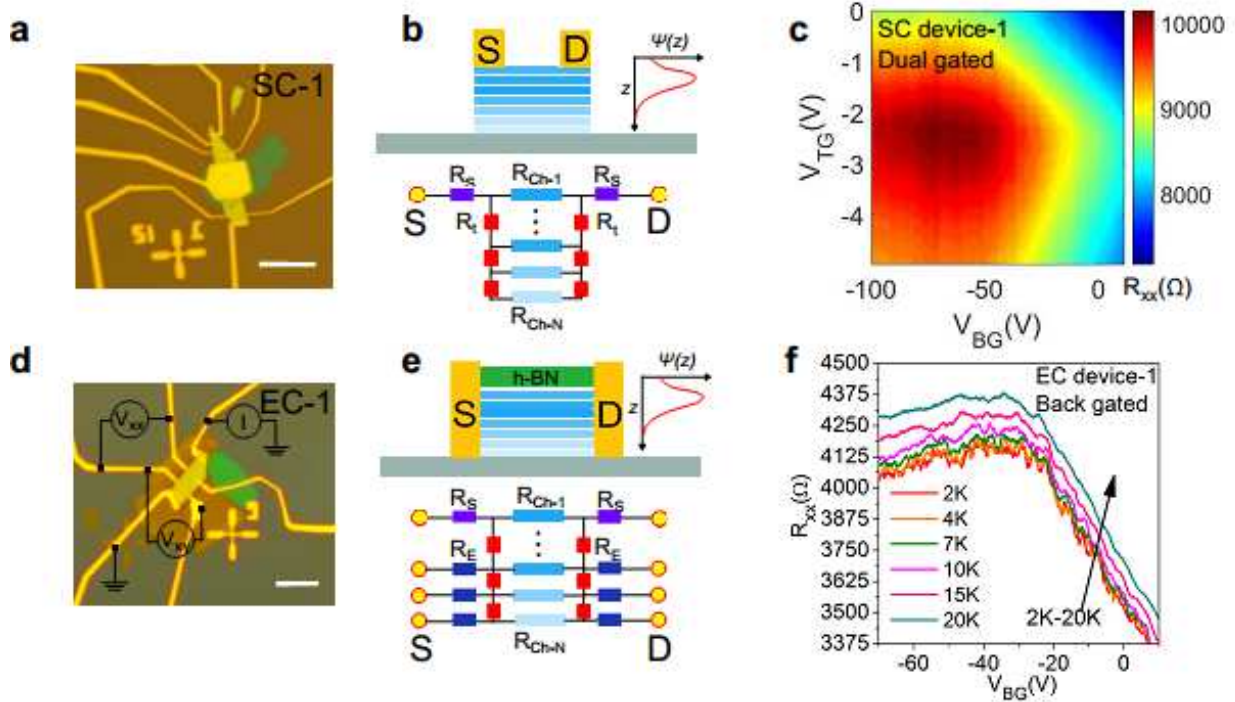}
\caption{(a) Optical image of a surface contacted device. (b) Schematic of surface-contacted measurement configuration and the corresponding resistance network model. R$_{ch,i}$, indicates the resistance of the i$^{th}$ layer from the top, which is exacted to vary according to the distribution of the topological surface state wave-function $\psi(z)$. (c) R$_{xx}$ as a function of \Vtg~and \Vbg~for a surface contacted device(SC-1). (d) Optical image of an edge contacted device (e) Schematic of edge-contacted measurement configuration and resistance model. Note that the edge-contact electrode directly connects with each layer through a resistance R$_{E}$. (f) R$_{xx}$ as a function of \Vbg for an edge contacted device(EC-1)}
\label{fig01}
\end{figure} 

Next, these stacks are patterned into either surface or edge contacted devices. For surface contacted(SC) devices, a single step lithography process is used to define electrical leads and a top-gate electrode as shown in Fig.~\ref{fig01}(a). The edge-contacted(EC) devices are fabricated using a two-step lithography process. In the first step, a poly-methyl-methacrylate (PMMA) layer is used to define the etch mask, followed by reactive ion-etching of the mesa to reveal the exposed h-bN/TI interface(see supplementary material section A for details). In the second lithography step, we define the contact pads in a Hall-bar geometry followed by e-beam evaporation of Cr/Au of thickness 10nm/100nm as shown in Fig.~\ref{fig01}(d). The heavily doped Si substrate acts as the back-gate electrode in all devices. Here, we present results five SC devices and three EC devices.

Both types of devices feature highly tunable electrical transport. As shown in Fig.~\ref{fig01}(c), surface contacted devices may be dual gated by using a combination of top and bottom gate voltages(\Vtg~and \Vbg~respectively). Charge neutrality(CNP) is obtained at \Vtg$\simeq$-2.5V and \Vbg$\simeq$-60V for device SC-1, and \Vbg$\simeq$-30V for the edge contacted EC-1 device(Fig.~\ref{fig01}(f)). While the exact position of the CNP varies from sample to sample in a small gate voltage window ($\Delta$\Vtg $\simeq$ 2-3V and $\Delta$\Vbg $\simeq$ 10-30V), surface charge carrier concentrations as low as 3-7$\times$10$^{12}$/cm$^2$ are routinely obtained(see supplementary material section E).  

While gating characteristics do not differ significantly between the SC and EC type of devices, the average sheet resistance of SC devices at the CNP, $\langle R_{sq,SC} \rangle=5899 \pm 1100 \Omega$ is significantly larger than EC devices,$\langle R_{sq,EC}\rangle=4350 \pm 320 \Omega$. The fundamental difference arises from the manner in which current is injected into the two device types. As shown in Fig.~\ref{fig01}(b), an electrical lead in a surface contact geometry injects charge carriers into the device only at sample surface where the lead terminates. This is due to the current crowding effect that inevitably occurs due to resistivity mismatch at metal-semiconductor interfaces~\cite{currentcrowding1,currentcrowding2} : charge current follows the path of least resistance, which in this case is the Cr/Au electrode until it reaches the BSTS surface where it is forced to eject into the topmost layer through a surface-contact resistance R$_S$. The layered nature of BSTS and the large bulk resistance allows this current to flow into the lower layers only through large inter-layer tunneling resistances R$_t$. Further, it is known that the topological surface state wave function is not completely localized at the top-most layer of the sample; rather it peaks a few layers into the bulk as shown in Fig.~\ref{fig01}(a). In fact, this delocalization effect is enhanced in the presence of disorder~\cite{disorderTI-1, disorderTI-2, disorderTI-3, disorderTI-4} or coupling with bulk states~\cite{bulk-surface, banerjee2017intermediate}. The surface contact geometry is therefore unable to completely access the TSS, leading to exaggerated values of sheet resistances. On the other hand, the edge contact geometry avoids both the current crowding and inter-layer tunneling resistance problems. As depicted in Fig.~\ref{fig01}(e), the edge contact directly connects with each of the layers in the BSTS flake through an edge-contact resistance $R_{E}$, and the TSS wave-function is uniformly probed. This results in a correct estimation of the channel resistance that is  lower than the resistances measured in the surface-contact geometry. Similar effects have been observed before in edge-contacted devices of multi-layer graphene and analyzed using resistor network models~\cite{EdgeContactRev1,EdgeContactRev2,edge-contact-graphene, edge-contact-graphene-1, edge-contact-graphene-2} .

\begin{figure}[!t]
\includegraphics[width=1.\linewidth]{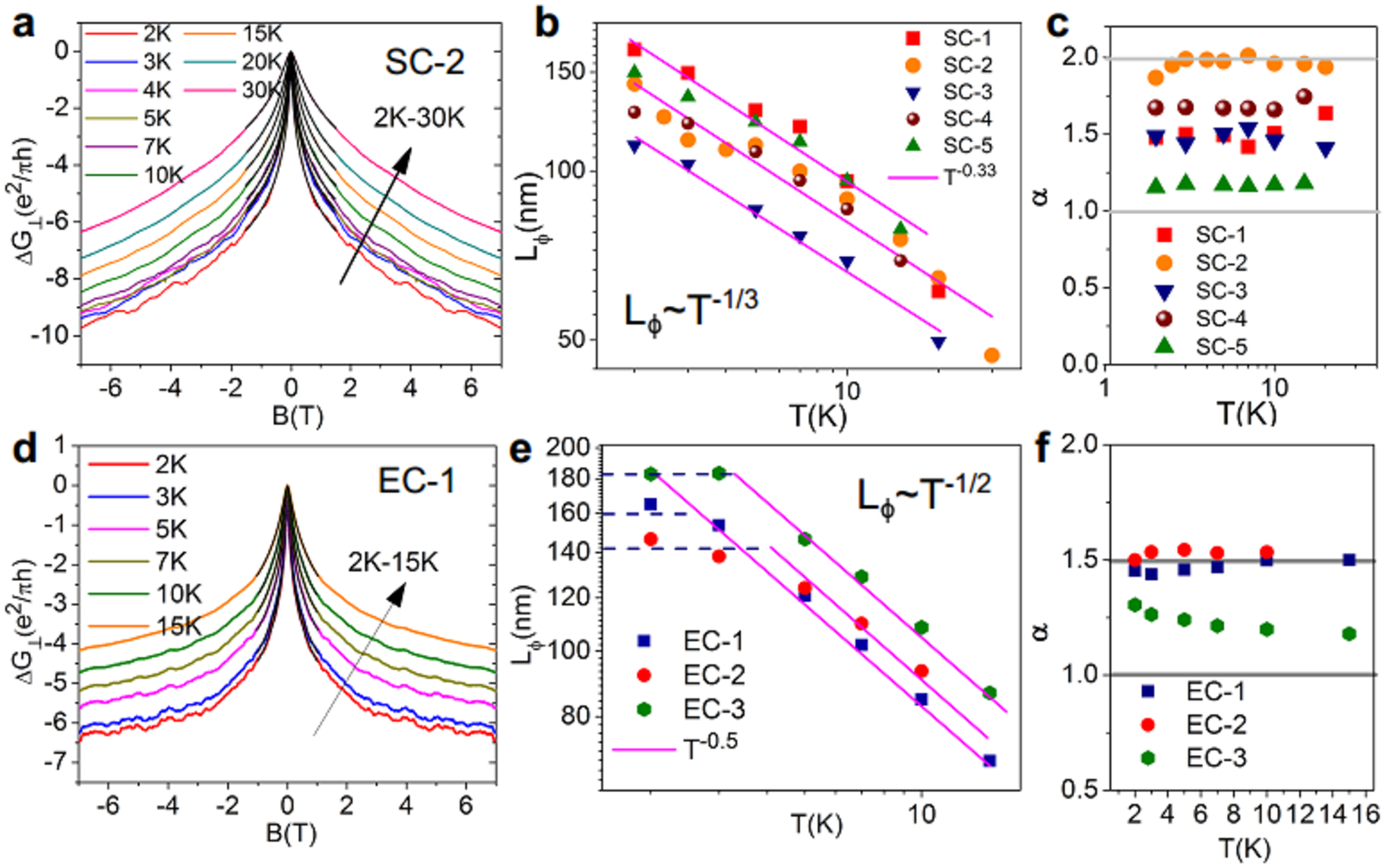}
\caption{(a) Magnetoconductance as a function of temperature for surface contacted sample SC-1. The low-field fits to the HLN formula are shown in solid black lines. (b) \lf and (c) \a as a function of temperature for devices surface contacted devices.  (d) Magnetoconductance and HLN fitting for edge contacted sample EC-1 (e) \lf and (f) \a as a function of temperature for edge contacted devices}
\label{fig02}
\end{figure}

\subsubsection{{\bf Phase coherent transport}}
Magnetoresistance measurements offer a powerful and reliable method to measure phase coherent transport properties. In the case of TIs, magnetoresistance due to weak anti-localization(WAL) can be used to independently estimate the electronic phase coherence length and the number of electronic channels that contribute to phase coherent transport. Such experiments carried out on different types of TI samples have revealed valuable topological physics~\cite{chen2010gate,checkelsky2011bulk,chen2011tunable,steinberg2011electrically, cha2012weak,Taskin,powerlaw2,BSTS2, gating5, lang2012competing, Banerjee16,banerjee2017intermediate,liao2017,inverseTI,WALAdditionalChannel}. The low field magnetoconductance in the WAL regime is described by the Hikami-Larkin-Nagaoka theory~\cite{HLN1, HLN2, HLN3,shen,WALcompetition} for corrections to conductance due to quantum interference effects. For a geometry where the sample is perpendicular to the magnetic field, the conductance correction is given as: 
\begin{equation}
\label{eq1} 
\Delta G_{xx}(B)=-\alpha \frac{e^2}{2\pi^2 \hbar} \left[\psi\left(\frac{1}{2} + \frac{\hbar}{4eL_\phi^2B}\right)-\ln \left(\frac{\hbar}{4eL_\phi^2B}\right)\right]
\end{equation}
 in the limit of strong spin-orbit scattering, where $\Delta G_{xx}$ is the change in sample conductivity, $B$ is the applied magnetic field, $L_\phi$ is the phase coherence length, and $\psi(x)$ denotes the digamma function. The leading constant $\alpha$ determines the number of {\it uncoupled} spin-orbit coupled channels, with a value of 0.5 for each such channel. 

 As shown in Fig.~\ref{fig02}(a) and (d) for a surface and edge contacted device respectively, the magneto-conductance(MC) drops sharply with increasing magnetic field forming the characteristic weak anti-localization(WAL) cusp. With increasing sample temperature, the WAL effect diminishes due to increasing channel decoherence. Fitting of the low-field MC to Eq.~\ref{eq1} yields \lf and \a, plotted as a function of temperature in Fig.~\ref{fig02}(b) and (c) for the SC devices and Fig.~\ref{fig02}(e) and (f) for the EC devices. For the SC devices(Fig.~\ref{fig02}(b)), \lf$\propto$T$^{-1/3}$ while for the EC devices(Fig.~\ref{fig02}(e)),  \lf$\propto$T$^{-1/2}$. The two different dephasing rates indicate two different carrier dephasing mechanisms. For dephasing to occur, a charge carrier must undergo inelastic collision and release/absorb energy. This loss(or gain) in energy causes the electronic wave-function to pick up an additional phase($\Delta \phi \propto \Delta E t/\hbar$) that eventually destroys its quantum coherence. In conventional two-dimensional electronic systems, the low temperature carrier dephasing is dominated by small-scale energy losses due to electron-electron interactions. This is the Nyquist dephasing regime~\cite{rammer,altshuler1981,altshuler1982}, that leads to a power-law dependence of \lf on temperature given by \lf$\propto$T$^{-1/2}$. This is exactly what we observe for the edge contacted devices. 
 
On the other hand,  \lf$\propto$T$^{-1/3}$ is indicative of dephasing due to variable range hopping(VRH)~\cite{VRH} When electrical transport is primarily of the VRH type, carriers hop between different energy levels provided by defect states within a band-width (-$\epsilon_0$, $\epsilon_0$) with $\epsilon_0\sim$kT. In this regime, every hop is caused by the absorption/emission of a phonon, leading to energy gain(or loss) for the carrier. Therefore, the phase coherence length is simply set by the hopping length $R_{hop}$~\cite{VRH2,VRH3,VRH1,VRH4}. In two dimensions, it can be easily shown(see supplementary material section G) that R$_{hop}\propto$T$^{-1/3}$. The observation of \lf$\propto$T$^{-1/3}$ in our surface contacted devices indicates VRH-type dephasing as the dominant decoherence mechanism in the SC-devices, while conventional Nyquist type e-e interaction survives in the EC-devices. We note here that similar VRH-type temperature dependence of \lf has been recently observed in (Bi$_{1-x}$Sb$_x$)$_2$Te$_3$ thin films when the samples were tuned into a bulk-insulating regime~\cite{liao2017}, and was attributed to an electronic coupling between surface states and bulk electron-hole puddles. 

Figures.~\ref{fig02}(c) and (f) show \a as a function of temperature for the SC and EC devices respectively. Contrary to most samples studied in literature so far, all devices(both EC and SC) show \a$\simeq1-2$. Here, $\alpha=1$ is contributed by the two decoupled top and bottom topological surface states. The additional contribution to \a$\simeq0-1$ arises from band-bending induced surface confined bulk carriers. These channels feature Rashba spin-splitting and have different carrier densities and mobilities, and can be identified using gate-dependent Hall effect measurements. For details we refer the reader to our previous work~\cite{Banerjee16}.  Here, we stress that such large values of \a reinforce our claim that the bulk of our samples is highly insulating: not only are the topological surface states decoupled from each other, but they are also decoupled from the Rashba surface states. While the Rashba surface states are inevitably present in most TI samples as measured through ARPES experiments~\cite{TIdecoration2, TIdecoration3, TI_2DEG}, they are hardly ever detected by electrical transport because of strong electrical coupling with the bulk states. The effect of bulk states can therefore be completely neglected in our studies making electrical transport primarily two-dimensional. This also justifies the use of the HLN model to analyze our experimental data, which is applicable only to two dimensional systems.
 
\begin{figure}[!t]
\includegraphics[width=1.1\linewidth]{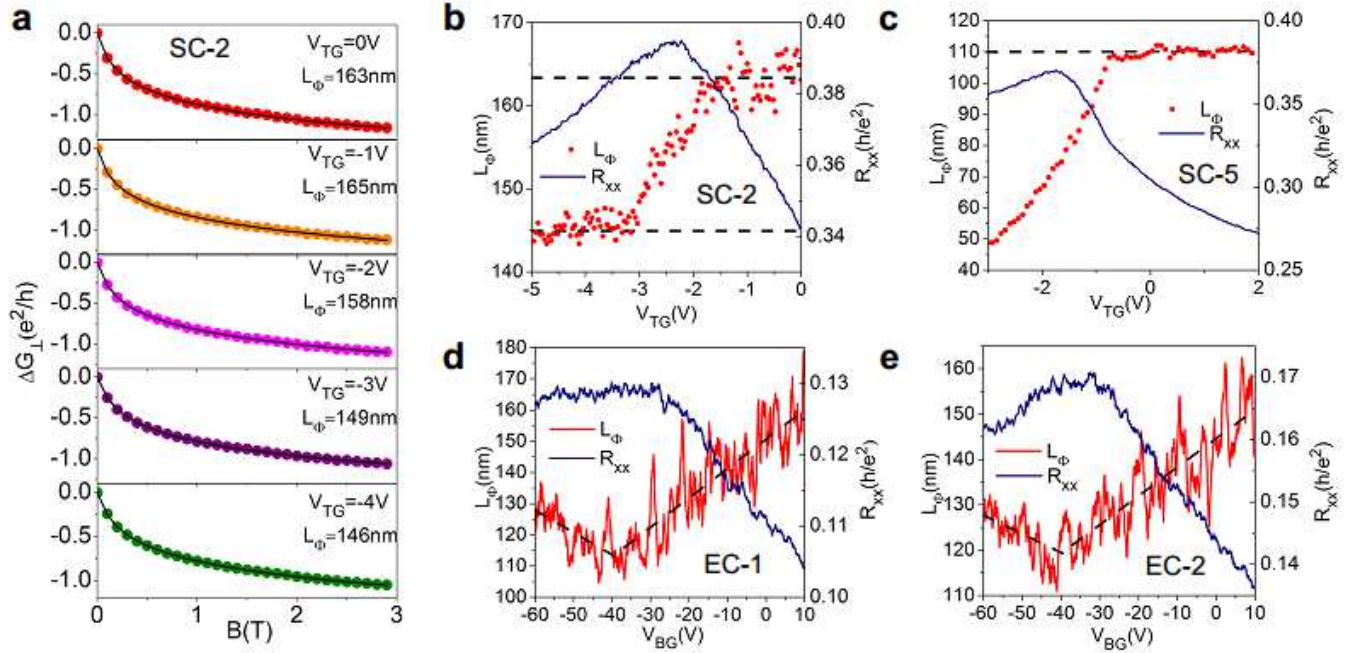}
\caption{(a) Magnetoconductance at different gate voltages for surface contacted sample SC-2 at T=2K. The low-field fits to the HLN formula are shown in solid black lines. (b) Phase coherence length \lf as a function of gate voltage(left y-axis) and corresponding sample resistance R$_{xx}$(right y-axis) for surface contacted samples SC-2 and (c) SC-5; and edge contacted samples (d) EC-1 and (e) EC-2.}
\label{fig03}
\end{figure} 

\subsubsection{{\bf Gate tunable phase coherence}}
Measurement of \lf as a function of gate voltage(at T=2K) reveals drastically different behavior for surface contacted versus edge contacted devices. Fig.~\ref{fig03}(b) and (c) depict \lf as a function of gate voltage for two surface contacted devices(SC-2 and SC-5). Qualitatively similar behavior is found in all surface contacted devices(see supplementary material section E). Away from charge neutrality, \lf becomes independent of the gate voltage, and remains saturated for a large gate voltage range(Fig.~\ref{fig02}(c)). Strong saturation of \lf is observed in both the n-type and p-type regions, however with different values. For example, in sample SC-2(Fig.~\ref{fig03}(b)), L$_{\phi,p}\sim$145nm in the p-type regime, but L$_{\phi,n}\sim$165nm in the n-type region. We consistently observe L$_{\phi,p}$ to be smaller than L$_{\phi,n}$ in all devices, with L$_{\phi,n}$$-$L$_{\phi,p}\simeq$20nm-60nm. In the intermediate gate voltage range, that is , close to charge neutrality, \lf sharply transitions between the two extremal values. In sharp contrast to this behavior, edge contacted devices show a cusp like dependence of \lf on gate voltage. At the CNP, \lf is minimum and it increases in either direction in a linear fashion. The change in \lf is $\sim$50nm in the measured gate-voltage window.

In the VRH regime, the hopping distance R$_{hop}$ is dependent on the dopant density of states(DOS) and sample temperature, but remains independent of the chemical potential as long as the density of dopant energy levels(N$_\mu$) within an energy window $\epsilon_0\sim kT$ of the chemical potential does not change(see supplementary material section G)~\cite{VRH}. In compensated semiconductors such as our material, the dopant density of states is not expected to vary with chemical potential, leading to a saturation of R$_{hop}$. The saturation of \lf in the surface contacted devices is a direct and rather remarkable consequence of this effect. When the chemical potential is tuned towards charge neutrality, the dominant dopant type changes from donor-like to acceptor-like. In compensated semiconductors, the donor dopant density(N$_d$) is close to the acceptor dopant density(N$_a$), but cannot be made to be exactly equal. This naturally leads to a transition of dopant DOS N$_\mu$ across the mid-gap energy level(E$_{mg}$), which roughly(though not exactly) corresponds to the Dirac point in our material as verified by ARPES measurements~\cite{lohani2017band}. Since R$_{hop}\propto N_\mu^{-1/3}$, \lf in the n-type region differs from that in the p-type region, but otherwise remains constant with the gate voltage.

On the other hand, the variation of \lf with gate voltage in edge contacted devices can be explained as a direct consequence of variation of e-e interactions with chemical potential. In diffusive conductors, disorder weakens screening and leads to so-called disorder enhanced e-e interactions~\cite{rammer,altshuler1981,altshuler1982}(see supplementary material section H), where the scattering $\tau_{ee}^{-1} \propto 1/\sigma_{xx}$, where conductivity $\sigma_{xx}=1/R_{xx}$. At the Dirac point, potential fluctuations due to e-e interactions are poorly screened because of low electrical conductivity, resulting in large e-e inelastic scattering and a corresponding minima in \lf. With increasing chemical potential, e-e interactions are weakened and \lf rises.

\begin{figure}[!t]
\includegraphics[width=0.8\linewidth]{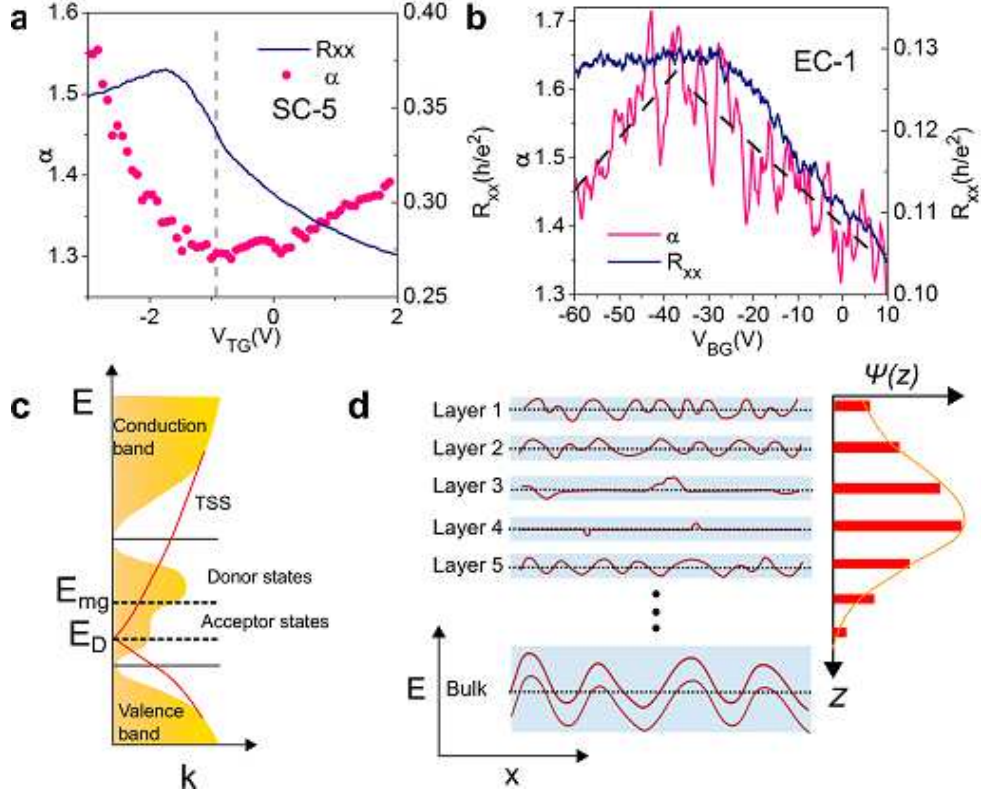}
\caption{(a) Magnitude of WAL effect \a as a function of gate voltage(left y-axis) and corresponding sample resistance R$_{xx}$(right y-axis) for surface contacted sample SC-5 and (b) edge contacted sample EC-1. (c) Schematic depicting the band-structure of a disordered topological insulator. Donor and acceptor like states populate in-gap energy levels and are separated by the mid-gap energy E$_{mg}$. The donor DOS is depicted larger than the acceptor DOS. The topological surface state(red solid line) co-exists with these states with the Dirac point E$_D$ lower than E$_{mg}$(d) Schematic depicting layer dependent disorder potentials due to variable screening by the topological surface state wave-function. Screening is maximum when the surface-state wave function has the largest amplitude. In the bulk, the impurity potentials are screened by bulk band-bending. Solid(red) curvy lines indicate indicate the disorder potential as a function of position. Dotted line(black) indicates the chemical potential. $\psi(z)$ represents the amplitude of the surface state wave-function in different layers.}
\label{fig04}
\end{figure} 

\subsubsection{{\bf Gate tunable suppression of weak-antilocalization}}
In the VRH regime, phase coherent transport occurs in a pseudo-diffusive regime. The Hikami-Larkin Naganoka theory describing weak localization(and anti-localization) effects assumes diffusive motion of electrons, and predicts the magnitude of weak anti-localization to be $\alpha e^2/(2\pi \hbar)$, where \a$=0.5$ is a fixed constant per independent electronic channel. This is however not valid in the pseudo-diffsuive regime, where it has been shown that $\alpha=0.5-2R/\pi$, where $R$ is the sample resistance in units of $e^2/h$~\cite{VRH_alpha,VRH_alpha1}. This suppression of \a is a consequence of pseudo-diffusive transport where quantum corrections to conductance need to take into account terms in higher orders of the dimensionless conductance $g=1/R$. Remarkably, in the VRH transport regime manifested in the SC devices, we observe a variation of \a with sample resistance that indicates pseudo-diffusive transport. As shown in Fig.~\ref{fig04}(a), \a shows a pronounced dip at the charge neutrality point when the sample resistance is highest, and increases in both directions away from charge neutrality. This suppression of WAL at the charge neutrality point is seen in all SC samples(see supplementary material section F). Such suppression of WAL has been observed before in TIs but only in the ultra-thin limit where topological surface states get tunnel coupled and purportedly undergo a topological-to-trivial phase transition, thereby allowing a suppression of WAL due to Anderson-type localization of carriers~\cite{Taskin, gating5, VRH_alpha1, lang2012competing}. However, our samples are in a fully topological regime featuring decoupled topological surface states. The suppression of WAL in our samples is therefore completely different from previous experimental observations, and is rather surprising given that topological surface states are generically immune to Anderson localization.

Contrasting this behavior with the edge-contacted devices, we observe an opposite trend. As shown in Fig.~\ref{fig04}(b), \a shows a peak instead of a dip at the charge neutrality point. Away from charge neutrality, \a decreases in a linear fashion. This effect has been observed previously in bulk-conducting topological insulators where the suppression of bulk-conductance by electrostatic gating decouples the opposite topological surface states, and leads to an increase in \a from 0.5 $\to$ 1~\cite{VRH_alpha1,gating5, chen2010gate, checkelsky2011bulk,steinberg2011electrically}. In our case, the topological surface states are already decoupled(since \a$>1.0$), but a pair of Rashba surface states remains quasi-coupled in the bulk~\cite{Banerjee16}, leading to an extra contribution $\Delta \alpha \simeq 0.5-1$. At charge neutrality, depletion of surface confined bulk carriers leads to a further decoupling of the two Rashba channels leading to an enhancement of WAL with $\Delta \alpha(V_G) \simeq 0.2$. 

\subsection{Discussion}
The surface contacted devices show three clear evidences is support of a variable range hopping type of transport: i) an anomalous temperature dependence of \lf$\propto T^{-1/3}$, ii) saturation of \lf  with the chemical potential and iii) suppression of WAL($\alpha$) as a function of increasing sample resistance. On the other hand, edge contacted devices manifest clear signature of e-e interaction dominated carrier dephasing: i) \lf$\propto T^{-1/2}$ and ii) linear variation of \lf with chemical potential; and an enhancement of WAL at the charge neutrality point due to channel-decoupling.
This is despite the fact that both types of samples were prepared under identical conditions in an inert environment and capped with a protective layer of h-BN. The difference therefore can originate only from the nature of the contacts to the two samples. 
 
The surface contact geometry injects(and detects) currents primarily into(from) the topmost layer of the sample. However, the top layer may not necessarily host the entire topological surface state wave-function. Although little is known experimentally  about the distribution of the topological surface state wave-function within the bulk of the material, indirect measurements of the TSS penetration depths using electrical transport indicate a value of $\sim$3nm for pristine samples, corresponding to three quintuple layers~\cite{Taskin}. However, previous experimental works by us~\cite{banerjee2017granular,banerjee2017intermediate} and several theoretical works~\cite{disorderTI-1, disorderTI-2, disorderTI-3, disorderTI-4} indicate that disorder can lead to strong delocalization of the surface-state wave-function, causing it to peak several layers into the bulk as shown in Fig.~\ref{fig04}(c). 

In a compensation doped TI, such an effect would lead to rather poor screening of ionized dopant potentials at the top surface, and better screening in layers where the surface-state local density of states peaks. The part of the wave-function residing on the top-most layer therefore suffers from strong scattering, leading to a VRH type of transport, while this disorder is quickly screened in deeper layers, where transport becomes diffusive. Measurement in an edge-contact geometry probes all the layers equally, and therefore senses diffusive electrical transport. On the other hand, the surface contact geometry only probes the topmost quintuple layer where transport is primarily of the VRH type. Of course, our qualitative understanding cannot capture the full physical picture: the distribution of the TSS wave-function is decided by the disorder configuration, which is in-turn determined by the screening due to the TSS wave-function. Such an analysis will require serious theoretical effort, and is beyond the scope of this work. Previous theoretical works have not taken this coupling into account, and therefore predicted that the effect of disorder is weakest at the surface due to strong screening from a perfectly surface-localized TSS~\cite{puddle4,skinner2012bulk,skinner2013effects}.

\subsection{Conclusions}
While compensation doping is key to obtaining true topological insulator behavior by reducing bulk carrier doping, its simultaneous effect on surface charge transport has gone unnoticed so far. This is perhaps motivated by an assumption of `immunity' of topological surface states to disorder. While it is true that Dirac fermion states cannot be strongly localized by disorder, electrical transport can still become non-diffusive due to resonant coupling with defect states~\cite{xu2017disorder}. We unveil the existence of such pseudo-diffusive transport in the top-layers of a  topological insulator by performing electron dephasing measurements that are much more sensitive to disorder than conventional transport or spectroscopic experiments. While the disorder strength is largest on the top-most quintuple layers, enhanced screening by the `bulk' part of the topological surface state wave-function restores diffusive transport in the inner layers of the material. Our work therefore provides a highly detailed picture of coherent surface charge transport in perhaps the most important TI material discovered so far. Further, to the best of our knowledge, such spatial dependence of electronic dephasing has not been observed before in any mesoscopic system, and is perhaps unique only to topological insulators. From a technical viewpoint, we show that that an edge-contacted scheme of measurement is mandatory for fully accessing topological surface state transport in TI devices. Apart from its application to TIs, our edge-vs-surface contacted measurement scheme may allow measurements of layer resolved electron dephasing in other 2D materials offering  unprecedented insights into electronic transport in this topical class of materials.

\subsection{Acknowledgement}
A.B. thanks MHRD, Govt. of India for support. A.S. thanks KVPY, Govt. of India for support. P.S.A.K. thanks Nanomission, DST, Govt. of India for support.

\bibliography{Electron_dephasing_TI_bib}

\end{document}